\documentstyle[twoside,fleqn,espcrc2,epsfig]{article}

\def\mr{\rm}	% {\mathrm}
\def\mbf{\bf}	% {\mathbf}
\def\emph#1{{\em #1\/}}

\def\veg#1{{\mbf #1}}
\def\vek#1{\mbox{\protect\boldmath $#1$}}
\def\vvek#1{\rlap{$\hspace*{0.05em}#1$}#1\hspace*{0.05em}}

\def\LQCD{\Lambda_{\mr QCD}}

\def\Order{{\mr O}}

\def\gsim{\mathop{\raisebox{-.4ex}{\rlap{$\sim$}} \raisebox{.4ex}{$>$}}}

\def\Order{{\mr O}}

\newcommand{\sixth}{\mbox{\small $\frac{1}{6}$}}
\newcommand{\third}{\mbox{\small $\frac{1}{3}$}}

\newcommand{\vdot}{\!\cdot\!}
\def\zahlen{{\sf Z\kern-0.4em Z}}

\def\dfrac#1#2{{\displaystyle \frac{#1}{#2}}}

\title{Binding Energies in Nonrelativistic Field Theories
\hfill {\normalsize hep-lat/9608139}
}

\author{Andreas S. Kronfeld\address{
Theoretical Physics Group, Fermi National Accelerator Laboratory,
Batavia, Illinois, USA}%
\thanks{Poster given at {\em Lattice '96}, 4--8 June 1996,
St.~Louis, Missouri}
\hfill {\normalsize FERMILAB-CONF-96/237-T}
}

\begin{document}

\begin{abstract}
Relativistic corrections communicate the binding energy of a bound 
state to its kinetic mass.
This mechanism is reviewed and used to explain anomalous results of 
Collins, Edwards, Heller, and Sloan (hep-lat/9512026), which compared 
rest and kinetic masses of heavy-light mesons and quarkonia.
\end{abstract}

\maketitle

\section{INTRODUCTION}
Last year Collins, Edwards, Heller, and Sloan~\cite{Col95} studied 
heavy Wilson quarks~\cite{Wil77} with the improved action of 
Sheikholeslami and Wohlert~\cite{She85}.
In lattice units the heavy-quark mass~$m_{Q}a$ typically exceeded 
unity, a regime in which the numerical results require a 
nonrelativistic interpretation~\cite{K&M93,KKM96} just as in 
NRQCD~\cite{Lep87,Lep92}.
Ref.~\cite{Col95} presents a test of the nonrelativistic 
interpretation that removes kinematic effects and focuses on a 
dynamical effect---the binding energy.
The results of this test were unexpectedly anomalous.
The aim of this paper is to explain why and to offer a remedy.

Let us begin with some basics to define notation.
As a function of momentum~$\vek{p}$ the energy of a state~$X$ can be 
written
\begin{equation}
	E_X(\vek{p}) = M_{1X} + \frac{\vek{p}^2}{2M_{2X}} - 
	\frac{(\vek{p}^2)^2}{8M^3_{4X}} + \cdots,
	\label{E-p}
\end{equation}
where the rest mass is defined by $M_1=E(\veg{0})$, and the kinetic mass 
is defined by 
\begin{equation}
	M_2 = \left(
		\frac{\partial^2E}{\partial p_i^2}
		\right)^{-1}_{\vvek{\scriptstyle p}=\vvek{\scriptstyle 0}}.
	\label{M2}
\end{equation}
Below the states can be quarks,~$Q$ and~$q$, and 
mesons~$\bar{Q}Q$, $\bar{Q}q$, and~$\bar{q}q$.
Usually~$Q$ is assumed heavier than~$q$.

On a relativistic mass shell 
\begin{equation}
	M_{1X}=M_{2X}=\cdots=m_X,
	\label{shell}
\end{equation}
In this paper the lower-case~$m_X$ denotes the exact, physical mass, 
whereas upper-case~$M_{iX}$ denote the result of a (possibly 
approximate) calculation.
In the mass-dependent renormalization of ref.~\cite{KKM96} it is 
possible to adjust the action's couplings to recover 
eq.~(\ref{shell}), to a specified accuracy.
With the Wilson or Sheikholeslami-Wohlert actions, however, 
perturbation theory shows that $M_{1Q}\neq M_{2Q}$ (except as 
$m_{Q}a\to 0$).
In nonrelativistic systems this is acceptable, provided one adjusts 
the bare mass until the \emph{kinetic} mass takes its physical 
value~\cite{K&M93,KKM96}, just as in ref.~\cite{Lep87,Lep92}.

The quark state makes sense at most in perturbation theory.
In a nonperturbative world, one would like to carry out the tuning 
with a bound state, e.g.~a meson $X=\bar{Q}q$ whose masses 
$M_{1\bar{Q}q}$, $M_{2\bar{Q}q}$, etc, can be computed with 
Monte Carlo methods.
Let us define the binding energy~$B$ by
\begin{equation}
	\begin{array}{r@{\;=\;}l}
		M_{1\bar{Q}q} & M_{1\bar{Q}} + M_{1q} + B_{1}, \\[0.7em]
		M_{2\bar{Q}q} & M_{2\bar{Q}} + M_{2q} + B_{2}.
	\end{array}
	\label{binding}
\end{equation}
To make a precise definition of the binding energies, one requires a 
precise definition of the quark masses in eq.~(\ref{binding}).
Here it is enough to take the rest and kinetic masses of the free 
theory ($g_0^2=0$) with the \emph{same} bare quark masses.
Computing bound-state splittings through rest masses makes sense only 
if~$B_1$ is the experimental binding energy, to sufficient accuracy.
Similarly, tuning the meson mass to the kinetic mass makes sense only 
if also~$B_2$ is the experimental binding energy, to sufficient accuracy.

As shown below, the anomalous result of ref.~\cite{Col95} is sensitive 
to $B_2-B_1$.
There are two keys to understanding it.
First, one must be careful about the qualifying phrase ``to sufficient 
accuracy'' in the preceding paragraph.
Second, one must to know how field theories communicate the binding 
energy to a bound state's kinetic mass.

Sect.~\ref{test} recalls the diagnostic test of ref.~\cite{Col95}.
Sect.~\ref{cutoff} assesses the cutoff effects of the binding energy 
in light-light and heavy-light mesons, and in quarkonium.
The mechanism for generating the ``kinetic binding energy'' is 
reviewed for a relativistic (continuum) gauge theory in 
sect.~\ref{Breit} and generalized to lattice gauge theory in 
sect.~\ref{lat}.
Sect.~\ref{conclusions} draws a few conclusions.

\section{THE TEST} 
\label{test}
Let us abbreviate $\delta M:=M_2-M_1$ and $\delta B:=B_2-B_1$.
Ref.~\cite{Col95} introduces
\begin{equation}
	I := \frac{2\delta M_{\bar{Q}q} -
		(\delta M_{\bar{Q}Q} + \delta M_{\bar{q}q})}{2M_{2\bar{Q}q}}.
	\label{I M}
\end{equation}
Comparison with eqs.~(\ref{binding}) shows that the quark masses drop 
out, leaving
\begin{equation}
	I  = \frac{2\delta B_{\bar{Q}q} -
		(\delta B_{\bar{Q}Q} + \delta B_{\bar{q}q})}{2M_{2\bar{Q}q}}.
	\label{I B}
\end{equation}
If the lattice action(s) of the quarks were sufficiently accurate, all 
$\delta B$s, and hence~$I$, would vanish.
($I$ vanishes trivially when $m_{\bar{Q}}=m_q$, even if $\delta 
B_{\bar{Q}q}\neq 0$.)

\begin{figure}[tbp]
 	\epsfxsize=0.46875\textwidth
	\epsfbox{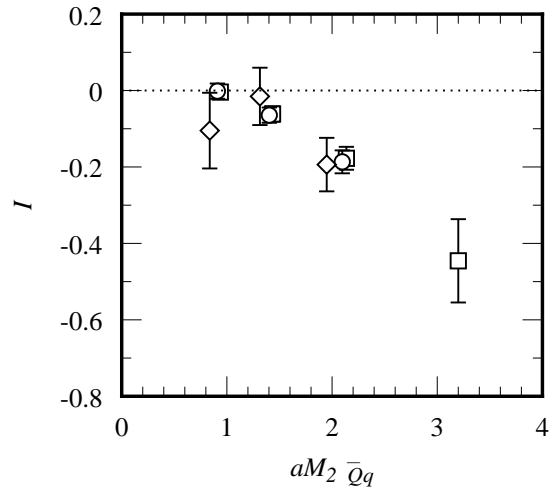} \vskip -0.8cm 
% 	\centerline{\includegraphics{sloan2.eps}}
	\caption[fig:sloan]{Plot of the binding-energy
        ``inconsistency''~$I$ vs.\ the heavy-light meson's kinetic mass
        $aM_{2\bar{Q}q}$, for~$m_Q\gg m_q$.
	Adapted from ref.~\cite{Col95}.} \vskip -14pt
	\protect\label{fig:sloan}
\end{figure}

The numerical results of ref.~\cite{Col95} are shown in 
fig.~\ref{fig:sloan}.
The ``inconsistency''~$I$ is negative, and~$|I|$ tends to increase 
with increasing~$m_Q$.
To explain both the sign and the magnitude, below I shall derive an 
expression for $\delta B$.

\section{CUTOFF EFFECTS ON $\delta B$}
\label{cutoff}
Before presenting the analytical result for $\delta B$, it is useful 
to anticipate the order of magnitude of $\delta B$ in each 
meson---light-light, heavy-light, and quarkonium.
On this basis it turns out that the quarkonium $\delta B_{\bar{Q}Q}$ 
dominates the numerator in eq.~(\ref{I B}).

\subsection{Light-light $\delta B_{\bar{q}q}$}
The binding energy is $\Order(\LQCD)$.
With the Sheikholeslami-Wohlert action, $B_1$ and $B_2$ both 
suffer from lattice artifacts of $\Order(\alpha^na\LQCD)$.
With the tree-level improvement of used by ref.~\cite{Col95},
$n=1$.
(With the Wilson action $n=0$.)
There is no reason for the artifacts to be identical, 
so $\delta B_{\bar{q}q}$ is $\Order(\alpha^na\LQCD^2)$.
This is numerically small, so $\delta B_{\bar{q}q}$ can be neglected 
below.

\subsection{Heavy-light $\delta B_{\bar{Q}q}$}
The binding energy is again $\Order(\LQCD)$.
The light quark suffers lattice artifacts as above, but, when 
$m_Q\gsim 1$---as in fig.~\ref{fig:sloan}, the heavy quark also 
suffers from (smaller) effects of $\Order(\LQCD^2 a/m_Q)$.
Again, even though there is no reason for artifacts in~$B_1$ 
and~$B_2$ to cancel, one sees that $\delta B_{\bar{Q}q}$ is 
numerically negligible.

\subsection{Quarkonium $\delta B_{\bar{Q}Q}$}
The binding energy is now $\Order(m_Qv^2)$, where~$v$ denotes the 
relative $\bar{Q}$-$Q$ velocity.
In this nonrelativistic system, the velocity is a pertinent estimator 
of cutoff effects~\cite{Lep92,KKM96}.
The rest mass is $\Order(v^0)$, so the action would need absolute 
accuracy of $\Order(v^2)$ to obtain relative accuracy of $\Order(v^2)$ 
in~$B_1$.
Both the Wilson and Sheikholeslami-Wohlert actions achieve this.
On the other hand, the kinetic mass multiplies an $\Order(v^2)$ 
effect, so the action would now need an absolute accuracy of 
$\Order(v^4)$ to obtain relative accuracy of $\Order(v^2)$ in~$B_2$.
Neither the Wilson nor the Sheikholeslami-Wohlert action achieves 
this~\cite{KKM96}; with either of them, one can only hope 
for~$\Order(v^0)$ relative accuracy in~$B_2$.
The error in~$B_2$, and hence in~$\delta B_{\bar{Q}Q}\neq0$, 
is~$\Order(m_Qv^2)$, which is significantly larger than the previous 
two estimates.

\section{BREIT EQUATION} 
\label{Breit}
For nonrelativistic systems the binding-energy discrepancy can be 
worked out quantitatively, following a textbook nonrelativistic 
expansion of QED~\cite{Ber71}.
This section verifies in a relativistically invariant theory that 
$B_2=B_1=B$.
The next section then turns to the lattice theories, which break 
relativistic invariance.
For convenience, these two section assume that even the ``light'' 
quark~$q$ is nonrelativistic.

At leading order the quark--anti-quark interaction arises from 
one-gluon exchange diagrams.
Evaluating these diagrams and developing the nonrelativistic 
expansion, one obtains a Hamiltonian $H=m_{\bar{Q}}+m_q+H_2+H_4$ for 
the quark--anti-quark system.
The leading nonrelativistic dynamics are given by
\begin{equation}
\begin{array}{r@{\,=\,}l}
H_2 & \dfrac{\vek{p}_{\bar{Q}}^2}{2m_{\bar{Q}}} +
      \dfrac{\vek{p}_q^2}{2m_q} + V(r) \\
    & \dfrac{\vek{P}^2}{2M_{\bar{Q}q}} + \dfrac{\vek{p}^2}{2\mu}   + V(r),
\end{array}
	\label{H2}
\end{equation}
where $V(r)=-C_F\alpha/r$; $\vek{r}=\vek{x}_{\bar{Q}}-\vek{x}_q$, 
$\vek{P}$ and~$\vek{p}$ are center-of mass coordinates and momentum; 
$\mu=(m^{-1}_{\bar{Q}}+m^{-1}_q)^{-1}$ is the reduced mass;
and $M_{\bar{Q}q}=m_{\bar{Q}}+m_q$.
The first relativistic corrections are
\begin{equation}
H_4 = - \frac{(\vek{p}_{\bar{Q}}^2)^2}{8m^3_{\bar{Q}}} - 
        \frac{(\vek{p}_q^2)^2}{8m^3_q} + V_2(\vek{r};\vek{p}_Q,\vek{p}_q),
	\label{H4}
\end{equation}
where the non-local potential $V_2$ is given by Breit's 
equation~\cite{Ber71}.
It takes the form
\begin{equation}
\begin{array}{l}
	V_2(\vek{r};\vek{p}_Q,\vek{p}_q) = 
	\mbox{const}\times\delta^{(3)}(\vek{r}) \\[0.7em]
	\hspace*{2.5em} +\,V(r)\left[1-
	    \dfrac{\vek{p}_{\bar{Q}}\vdot\vek{p}_q +
	           r^{-2}r_ir_j p_{\bar{Q}i}p_{qj}}{2M_{\bar{Q}q}\mu}
	    \right] \\[1.0em]
	\hspace*{2.5em} +\;\mbox{spin-dependent}.
\end{array}	 
	\label{V2}
\end{equation}
The spin-dependent terms and the terms proportional to 
$\delta^{(3)}(\vek{r})$ are not important here.
Full details are given in \S\S~83--84 of ref.~\cite{Ber71}.
Together with the $(\vek{p}^2)^2$ terms in~$H_4$, the exhibited part 
of~$V_2$ is responsible for modifying the bound-state kinetic mass 
from $M_{\bar{Q}q}=m_{\bar{Q}}+m_q$ to 
$m_{\bar{Q}q}=m_{\bar{Q}}+m_q+B$ (as required by Lorentz invariance).

To proceed one must re-write~$H_4$ in center-of-mass momenta and 
collect terms quadratic in the total bound-state momentum~$\vek{P}$.
In the bound state, combinations of the internal momentum~$\vek{p}$ 
and relative coordinate~$\vek{r}$ can be replaced by expectation 
values.
Collecting all terms, the bound-state kinetic energy becomes
\begin{equation}
\begin{array}{r@{\;}l}
	\dfrac{\vek{P}^2}{2m_{\bar{Q}q}}:= &
	\dfrac{\vek{P}^2}{2M_{\bar{Q}q}}
	\left(1-\dfrac{\langle T+V\rangle}{M_{\bar{Q}q}}\right) \\[1.0em]
	+ &
	\dfrac{P_iP_j}{2M^2_{\bar{Q}q}}
	\left\langle r_i\nabla_jV-p_ip_j/\mu\right\rangle,
\end{array}
	\label{Mm}
\end{equation}
where $T=\vek{p}^2/2\mu$ is the internal kinetic energy.
By the virial theorem the second line vanishes.
Thus, to consistent order in $\vek{p}/M$ the leading relativistic 
corrections~$H_4$ generates the right binding energy 
$B_2=\langle T+V\rangle=:B$ for the bound-state kinetic mass.

More generally, higher-order relativistic effects trickle down to 
bound-state properties as follows: the correction of~$\Order(v^\ell)$ 
provides the $\Order(v^{\ell-k})$ contribution to bound-state 
properties of~$\Order(v^k)$.

\section{LATTICE GENERALIZATION}
\label{lat}
On a hypercubic lattice there can be two corrections to the kinetic 
energy
\begin{equation}
	E(\vek{p}) = \cdots
	- \frac{(\vek{p}^2)^2}{8M^3_4}
	- \sixth w_4 a^3\sum_i p_i^4 + \cdots,
	\label{p^4}
\end{equation}
for each of $\vek{p}=\vek{p}_{\bar{Q}}$, $\vek{p}_q$.
Here
\begin{equation}
	M_4:= - \left(
	    \frac{\partial^4E}{\partial p_i^2\partial p_j^2}
	    \right)^{-1/3}_{\vvek{\scriptstyle p}=\vvek{\scriptstyle 0}},
        \quad i\neq j
	\label{M4}
\end{equation}
and
\begin{equation}
	w_4:= - \frac{1}{4} \left.\frac{\partial^4E}{\partial p_i^4}
	      \right|_{\vvek{\scriptstyle p}=\vvek{\scriptstyle 0}}
	      - \frac{3}{4M_4^3}.
	\label{w4}
\end{equation}
Unless the action has been improved \emph{further} than the 
Sheikholeslami-Wohlert action, $M_4\neq M_2$ and $w_4\neq 0$, 
cf.~Appendix~A of ref.~\cite{KKM96}.
These lattice artifacts filter through to~$B_2$---just as 
above---through the terms proportional to 
$P_iP_j\langle p_ip_j\rangle$.

The spatial gluon generates the spin-independent contribution 
proportional to $V(r)$ in eq.~(\ref{V2}); on the lattice the nuance is 
that the kinetic mass appears in the bracket.
On the other hand, the temporal gluon generates more complicated 
terms, but they either depend on spin or are proportional 
to~$\delta^{(3)}(\vek{r})$.
So to work out an expression for~$B_2$, it is enough to maintain 
eq.~(\ref{V2}), but with masses~$M_{2\bar{Q}q}$ and~$\mu_2$ built
from~$M_{2\bar{Q}}$ and~$M_{2q}$.

The calculation of the binding energy difference~$\delta B$ follows the
steps leading to eq.~(\ref{Mm}).
One finds an expression that is too cumbersome to present here.
In an S~wave, however, it can be simplified because
$\langle p_ip_j\rangle=\third\delta_{ij}\langle p^2\rangle$.
Then
\begin{equation}
\begin{array}{r@{}l}
\!\!
	\dfrac{\delta B}{\langle T\rangle} = 
		\third \left\{\rule{0.0em}{2.0em}\right. &
		5\left[\mu_2\left(\dfrac{M^2_{2\bar{Q}}}{M^3_{4\bar{Q}}} +
		\dfrac{M^2_{2q}}{M^3_{4q}}\right) -1 \right] \\[1.5em]
	+ &\,4 a^2\mu_2( M^2_{2\bar{Q}} w_{4\bar{Q}} + M^2_{2q} w_{4q})
	\left.\rule{0.0em}{2.0em}\right\}.
\end{array}
	\label{delta B}
\end{equation}
This is the main new result of this paper.
Note that, as one would have anticipated, the expression vanishes 
when $w_{4X}=0$ and $M_{4X}=M_{2X}$.

With an estimate of $\langle T\rangle$ from potential 
models~\cite{Kha96} and the lattice masses of the right-most point in 
fig.~\ref{fig:sloan}, I find
\begin{equation}
	I\approx-\frac{\delta B_{\bar{Q}Q}}{2M_{\bar{Q}q}}\approx -0.5.	
	\label{guess}
\end{equation}
The agreement with the Monte Carlo results is surprisingly good.

\section{CONCLUSIONS}
\label{conclusions}
The origin of the anomaly observed in ref.~\cite{Col95} is the usage 
of an action accurate only to~$\Order(v^2)$.
Thus the relative error in the binding energy~$B_2$ of the bound-state 
kinetic mass is of order $m_Qv^2/m_Qv^2=1$.
Meanwhile, the usual binding energy~$B_1$ is indeed valid to 
leading order in~$v^2$.
The test quantity~$I$ cleverly isolates $B_2-B_1$, and thus exposes 
an inconsistency of~$\Order(1)$.

By examining how (approximately) relativistic field theories 
generate~$B_2$, this paper explains the results found last 
year~\cite{Col95}. 
Moreover, the analysis makes the remedy plain:
the anomaly is not expected to appear if quarkonium properties are 
computed with an action improved through $\Order(v^4)$ (or higher).
In particular, one requires $M_4=M_2$ and $w_4=0$.

Most published applications of ref.~\cite{Lep87} use a sufficiently 
accurate action~\cite{Lep92}.
Ref.~\cite{Dav94} even remarks that $\Order(v^4)$ accuracy is 
essential for a consistent determination of the $b$-quark mass from 
the $\Upsilon$ spectrum.
For four-component fermions the details required for~$\Order(v^4)$ 
accuracy in quarkonium have appeared more recently~\cite{KKM96}.

\section*{ACKNOWLEDGEMENTS}
I would like to thank Sara Collins and John Sloan for bringing their
results to my attention and for insisting that I explain it.
Fermilab is operated by Universities Research Association, Inc.,
for the U.S. Department of Energy.


\begin{thebibliography}{99}
\bibitem{Col95} % the ``Sloan'' anomaly
S. Collins, R. Edwards, U. Heller, and J. Sloan, hep-lat/9512026,
Nucl. Phys. B Proc. Suppl. {\bf 47} (1996) 455.
\bibitem{Wil77} % Cargese lectures (solve doubling problem)
K.G. Wilson, in {\em New Phenomena in Subnuclear Physics},
A. Zichichi (ed.), Plenum, New York, 1977.
\bibitem{She85}
B. Sheikholeslami and R. Wohlert, Nucl. Phys. {\bf B259} (1985) 572.
\bibitem{K&M93}
P.B. Mackenzie, Nucl. Phys. B Proc. Suppl. {\bf 30} (1993)  35;\\
A.S. Kronfeld,  Nucl. Phys. B Proc. Suppl. {\bf 30} (1993) 445, 
{\bf 42} (1995) 415.
\bibitem{KKM96}
A.X. El-Khadra, A.S. Kronfeld, and P.B. Mackenzie, hep-lat/9604004, 
FERMILAB-PUB-96/074-T, ILL-TH-96-1.
\bibitem{Lep87} % NRQCD
G.P. Lepage and B.A. Thacker,
Nucl. Phys. B Proc. Suppl. {\bf 4} (1987) 199;\\
B.A. Thacker and G.P. Lepage, Phys. Rev. {\bf D43} (1991) 196.
\bibitem{Lep92} % Improved NR QCD for Heavy Quark Physics
G.P Lepage, L. Magnea, C. Nakhleh, U. Magnea, and K. Hornbostel,
Phys. Rev. {\bf D46} (1992) 4052.
\bibitem{Ber71}
V.I. Berestetskii, E.M. Lifshitz, and L.P. Pitaevskii,
\emph{Relativistic Quantum Theory}, Pergamon, Oxford, 1971.
\bibitem{Kha96}
A.X. El-Khadra, private communication.
\bibitem{Dav94}
C.T.H. Davies et al, Phys. Rev. Lett. {\bf 73} (1994) 2654.
\end{thebibliography}
\end{document}